\documentclass[12pt]{article}
\usepackage{amsmath}
\usepackage{amssymb}
\usepackage{citesort}
\usepackage{graphicx}
\usepackage[nomarkers]{endfloat}
\usepackage{pstricks}
\begin{document}
\setcounter{page}{0}
\thispagestyle{empty}
\begin{flushright}
\end{flushright}
\vspace*{2.5cm}
\begin{center}
{\large\bf On the Primordial Magnetic Field  \\
  from Domain Walls}
\end{center}

\vspace*{2cm}

\renewcommand{\thefootnote}{\fnsymbol{footnote}}

\begin{center}
{ P. Cea$^{1,2,}$\protect\footnote{Electronic address: {\tt
Cea@ba.infn.it}},
and
L. Tedesco$^{1,2,}$\protect\footnote{Electronic address: {\tt
luigi.tedesco@ba.infn.it}} \\[0.5cm] $^1${\em Dipartimento di Fisica,
Universit\`a di Bari, I-70126 Bari, Italy}\\[0.3cm] $^2${\em INFN
- Sezione di Bari, I-70126 Bari, Italy} }
\end{center}

\vspace*{0.5cm}

\begin{center}
{
November, 2002 }
\end{center}

\vspace*{1.0cm}

\renewcommand{\abstractname}{\normalsize Abstract}
\begin{abstract}
In this paper we discuss once more
the zero mode contribution
to the vacuum energy density.  We  show that a careful  treatment
of the zero modes
leads to the conclusion that domain walls may be ferromagnetic, and
could  generate a  magnetic field of cosmological interest.
\end{abstract}
\vfill
\newpage

In Ref.~\cite{Cea:1999} we suggested that the spontaneous
generation of uniform magnetic condensate in
$QED_3$~\cite{Cea:1985},~\cite{Hosotani:93},~\cite{Cea:2000} could
give rise to ferromagnetic domain walls at the electroweak phase
transition. Moreover, we suggested that these domain walls
generate a magnetic field at the electroweak scale which can be
relevant for the generation of the {\it primordial magnetic field}
\cite{DOLGOV}. In this paper we focus on some relevant points of
our technique with respect to other approaches. In particular in
Ref.~\cite{Voloshin:2000}, for example, the author points out that
massive $(2+1)$ dimensional fermions bound to a domain wall behave
diamagnetically rather than ferromagnetically. This  delicate
question concernes  the magnetic field contribution to the
fermionic vacuum free energy discussed for the first time in
Ref.~\cite{Cea:1985}.
\\
The aim of the present paper is twofold. First, we show  that
the results of ~\cite{Voloshin:2000} derive from an incorrect
treatment of the fermionic zero mode contribution to the zero
temperature vacuum energy density. Secondly, we show that the
finite temperature calculation of Ref.~\cite{Voloshin:2000} agrees
with our previous findings~\cite{Cea:2000},~\cite{Cea:1998}.
Moreover, if one takes into account the correct definition of an
Abelian magnetic field in presence of a varying scalar field
condensate, then it turns out that the classical magnetic energy
is proportional to the area of the wall. This last result supports
our previous proposal in Ref.~\cite{Cea:1999}. \\
The crucial point of the question is  the zero temperature vacuum energy
density due to
massive planar fermions in presence of a constant magnetic field:
\begin{equation}
\label{Eq.1}
E_{\rm vac}= - {e \, B \frac {2} {\pi}} \,\sum_{n=0}^\infty
\sqrt{2 \, e \, B \, n +m^2}~.
\end{equation}
The infinite sum in Eq.~(\ref{Eq.1}) needs to be regularized.
Among the possible gauge-invariant choices, we employed the
Schwinger proper-time regularization scheme. However, any gauge
invariant regularization gives physically sensible results.
Following Ref.~\cite{Voloshin:2000} we regularize the sum by means
of the gauge invariant cut-off $\exp(-\epsilon \, E_n^2)$, where
$\epsilon$ is the regulator parameter and $E_n$ is the energy of
the levels. We get:
\begin{equation}
\label{Eq.2}
E_{\rm vac}^{(r)}(B)= -  \frac {e \, B}  {2 \pi}
\,\sum_{n=0}^\infty \sqrt{2 \, e \, B \, n +m^2} \, \exp(-\epsilon
\, 2 \, e \, B \, n - \epsilon \, m^2)~.
\end{equation}
To evaluate the sum in Eq.~(\ref{Eq.2}), following Ref. \cite {Voloshin:2000}
we use the
Poisson's summation formula:
\begin{equation}
\label{Eq.3}
\sum_{n=0}^\infty \, f(n) = \int_{\delta}^\infty \, f(x) \,
\sum_{n=-\infty}^\infty \, \delta(x-n) \, dx =
\sum_{k=-\infty}^\infty \, \int_{\delta}^\infty \, f(x) \, \exp( 2
\, \pi \, i \, k \, x) \, dx~~,
\end{equation}
where $\delta$ is such that $-1 < \delta <0$. Now, observing that
the summand in Eq.~(\ref{Eq.2}) is non singular at $n=0$, the author of
Ref. \cite {Voloshin:2000}

assumes that one can set $\delta=0$. However it is easy to see
that this last assumption is not valid. Indeed, if we consider the
$n=0$ term in Eq.~(\ref{Eq.3}) :
\begin{equation}
\label{Eq.4}
I_\delta \equiv \int_{\delta}^\infty \, f(x) \, \delta(x) \, dx \;
,
\end{equation}
we see that $\lim_{\delta \rightarrow 0^-} I_\delta =
f(0)$, while
$\lim_{\delta \rightarrow 0^+} I_\delta = 0$. It is now evident
that the limit $\delta \rightarrow 0$ is not harmless. Indeed, the
Ref. \cite {Voloshin:2000} procedure is equivalent to ignore the zero
mode contribution. \\
The correct procedure can be obtained as follows. First one must
isolate the zero mode contribution and, then, one can apply the
Poisson's summation formula to the $n \geq 1$ modes. In this way,
in the weak magnetic field region we get for the vacuum energy
density:
\begin{eqnarray}
\label{Eq.5}
 E_{\rm vac}(B)-E_{\rm vac}(0)
& = &
 -  \frac {eB} {4 \pi} |m| + \frac {e^2 \, B^2}
 { 2 \, \pi^3 \, |m|} \sum_{p=0}^\infty \frac {(-1)^p \, \zeta(2 \, p +2)
\, \Gamma(\frac {3} {2}) } { \Gamma (\frac {1}  {2} - 2 \, p)} \,
\left ( \frac {e \, B} { \pi \, m^2} \right )^{2p}
 \nonumber \\
 &  &
  = - \frac {eB} {4 \pi} |m| + \frac {e^2 \, B^2} { 24 \, \pi \, |m|} +
\ldots ~.
\end{eqnarray}
This last equation agrees with the results obtained in
Ref.~\cite{Cea:1985}, and differs from Eq.~(14) of
Ref.~\cite{Voloshin:2000} in the negative term, linear in the magnetic
field. Obviously, it is the linear term which gives rise to
the spontaneous magnetic condensation in $QED_3$.
\\
Equation~(\ref{Eq.5}) shows that the  results reported in
Ref.~\cite{Voloshin:2000}, corrected to include the zero mode
contributions,  fully support the spontaneous generation of a
uniform magnetic condensate in $QED_3$ with massive fermions.
Moreover, it is worthwhile to stress that even the free energy
higher temperature study of Ref.~\cite{Voloshin:2000} is in
agreement with the one reported in
Ref.~\cite{Cea:2000},~\cite{Cea:1998}. Indeed, if one takes into
account the missing linear term in the zero temperature
energy density, it easy to see that Eq.(18) of
Ref.~\cite{Voloshin:2000} is in accordance with our previous
conclusion that the thermal corrections, even at infinite
temperature, do not modify the spontaneous generation of the
magnetic condensate. \\
Let us now comment on the  claim in Ref. \cite {Voloshin:2000},
that the domain walls
cannot be a source of the primordial magnetic field. This
statement  is based  on the argument that
the classical energy of the induced magnetic field is proportional
to the volume of the box, while the contribution due to the
fermion modes localized on the wall scales with the area of the
wall. In this way one could obtain for the
total energy of the system
\begin{equation}
\label{Eq6}
 E(B)=L^3 \, \frac {B^2 } {2} + L^2 \, f(B) \;,
\end{equation}
where $L$ is the linear size of the system. It turns out that the
above equation forgets completely the non Abelian nature of the
induced magnetic field. Indeed, in the case of varying scalar
field condensate the correct definition of Abelian electromagnetic
field is given by the t'Hooft's Abelian
projection~\cite{'tHooft:1974}. Taking into account that in our
model the Abelian part of the Abelian projected magnetic field is
induced by the fermionic modes localized on the wall, it is easy
to see that the Abelian projected magnetic field vanishes in the
regions where the scalar condensate is constant. As a matter of
fact, one can think of our field configuration in terms of a sheet
of Abelian magnetic monopoles of constant surface density
localized on the surface of scalar condensate zeroes. Such
monopole sheet  gives rise to an almost constant magnetic field
perpendicular to sheet which extends over a distance of order of
the wall thickness $\Delta$. Indeed it is easy to see that
the appropriate
expression for the Abelian magnetic field can be different from
zero only in the regions where the scalar condensate varies. It is
evident, now, that the Abelian projected magnetic field increases
the energy of the system by the "classical energy" $\sim L^2
\Delta \frac {B^2} {2}$ instead of $L^3 \frac {B^2} {2}$ in
Eq.~(\ref {Eq6}). Thus Eq.~(\ref{Eq6}) is replaced by:
\begin{equation}
\label{Eq.7}
 E(B)=L^2 \, \Delta \, \frac {B^2}  { 2} + L^2 \, f(B) \;,
\end{equation}
which led to the conclusions of Ref.~\cite{Cea:1999}. Recently, in
Ref. \cite {volo2} it is pointed out that due to $C$ parity
arguments it cannot be spontaneous generation of magnetic field in
$(3+1)$-dimensions. Our previous discussion shows that the
criticism of Ref. \cite {volo2} does not apply. Indeed, our
peculiar magnetic field can be generated due to the non Abelian
nature of the electroweak  gauge groups where charge conjugation
is no longer a symmetry.
\\
In conclusion, it is worthwhile to stress that in the realistic
case where the domain wall interacts with the plasma, the magnetic
field penetrates into the plasma over a distance of the order of
the penetration length $\lambda$ which is about an order of
magnitude greater than $\Delta $. It follows that the estimate in
Ref.~\cite{Cea:1999} of the induced magnetic field at the
electroweak scale $ B^* \, \simeq \, 5 \, 10^{24} \, Gauss$ is
reduced by a factor $\sqrt{\frac {\Delta} {\lambda}} \, \sim \,
0.3$  which, however, is still of cosmological interest.
However, it is worthwhile to stress that a complete treatment
of the problem requires the discussion of the generation and evolution
of the domain walls and their interaction with the surrounding plasma.
%
%
\vspace*{4.5cm}
\end{document}